\documentstyle[prd,aps,preprint]{revtex}
\input{epsf}
\begin{document}


\title{Quench Induced Vortices in the \\Symmetry Broken Phase of Liquid $^4$He}
\author{A.J. Gill$^{1,2,3,4}$ and T.W.B. Kibble$^{1,2}$}
\address{$^1$ Blackett Laboratory, Imperial College, South Kensington,
London SW7 2BZ, U.K. \\ $^2$ Isaac Newton Institute For Mathematical
Sciences,\\ 20 Clarkson Road, Cambridge CB3 0EH, U.K. \\
$^3$ Theoretical Division MS B288, Los Alamos National Laboratory, \\
Los Alamos, NM 87545, U.S.A.\\ and \\
$^4$ Low Temperature Laboratory,
Helsinki University of Technology,
02150 Espoo,
Finland }
\date{\today}

\maketitle

\medskip
\begin{center}
LAUR 95 - 4505 \\ Imperial/TP/95-96/29
\end{center}
\centerline{PACS: 67.40.V, 11.10.W, 05.70.F}

\vskip 2cm
\begin{abstract}
Motivated by the study of cosmological phase transitions, our understanding
of the formation of topological defects during spontaneous symmetry-breaking
and the associated non-equilibrium field theory has recently changed.
Experiments have been performed in superfluid $^4$He to test the new ideas
involved. In particular, it has been observed that a vortex density is seen
immediately after pressure quenches from just below the $\lambda$
transition.  We discuss possible interpretations of these vortices, conclude
they are consistent with our ideas of vortex formation and propose a
modification of the original experiments.
\end{abstract}

\vskip 1cm

\section{Introduction And Overview}

The formation of topological defects is generic to many physical systems, for
example superfluid $^4$He and $^3$He, superconductors and certain
grand-unified theories which may be relevant to the early universe. The
dynamics of vortex formation in all of these systems is intrinsically
{\it non}-equilibrium due to the symmetry-breaking phase transition
involved but despite their variety they can all be described by
some form of field theory and their physical behaviour is very similar.
Clearly there is some general feature of the behaviour of quantum fields out
of equilibrium which is causing the universality of this behaviour.
Such systems therefore provide a good arena
for the study of non-equilibrium field theory, both experimentally and
theoretically, since the defect density formed provides a good
experimental test of the underlying field theory.

Here we are concerned with an experiment designed to test our ideas of
what determines the defect density initially formed immediately after
{\it second order} symmetry breaking phase transitions.
The {\it mechanism}, sometimes called the Kibble mechanism, whereby
topological defects, and vortices in particular,
are formed \cite{twbk} is understood
and believed to be correct. It is not entirely certain, however, what
sets the scale in this mechanism and thus
determines the defect density immediately after the transition, although
various scenarios have been proposed.
It was originally thought that the defect density was determined when the
system had cooled sufficiently far below the transition that thermal
fluctuations were no longer important
\cite{ginzburg}. In other words it was thought
that the string network was `frozen in' at a temperature, the Ginzburg
temperature, when there was no longer sufficient thermal energy around for
the field to fluctuate out of topologically stable configurations like
strings. A criterion based on temperature cannot be wholly reliable
because non-equilibrium physics is involved in the phase transition.
An alternative, called the Zurek
scenario \cite{whz}, was suggested in which, due to the critical slowing down
involved  in second order transitions, the field is unable to keep up with
the quench  during an adiabatic period of evolution around the transition.
The  length scale of the initial vortex network is then the correlation
length of the field when it first comes back into equilibrium below the
transition which in this scenario is equal to the correlation length when the
field went out of  equilibrium above the transition.

An experiment, which we will refer to as the Lancaster experiment,
designed to distinguish between these two alternative scenarios has been
performed in superfluid $^4$He \cite{pvem}.
Preliminary results indicate the Zurek scenario to be correct, at least
within its domain of applicability. There are, however, still some
controversies and unresolved issues concerning whether or not the defect
density might not perhaps have been formed by
some means considered more conventional in condensed matter physics.
In particular there is an unexplained observation of vorticity after
quenches from below the superfluid or $\lambda$ transition. One of our
aims is
to explain this and thereby clarify the interpretation of the experiment.
The experiment will then give us insight into the non-equilibrium
behaviour of scalar fields during symmetry breaking phase transitions.

\section{The Formation Of Vortices}

Here we are interested solely in vortex production during second order
transitions which break a $U(1)$ symmetry and produce vortices, as in
superfluid $^4$He \cite{rpf}. While there is no particularly good field
theoretical model for superfluid $^4$He, to perform calculations we later
assume a particular model \cite{wiegel}
which exhibits all the generic features of a $U(1)$ symmetry-breaking
phase-transition in which vortices are formed. We will also limit ourselves to
the case of phase transitions sufficiently rapid that the Zurek scenario can
be expected to be reasonably accurate.

The mechanism by which vortices are formed during such transitions is well
understood \cite{twbk}. In the
symmetry-broken phase, above the phase-transition, the order parameter is
zero. As the transition is driven, by somehow removing energy from the system,
the field begins to notice the central hill of the Mexican hat
symmetry-breaking potential. Thus, the field tends to assume non-zero values
and to increase in magnitude until it reaches its vacuum expectation value.
There is no reason, however, for it to acquire the same phase everywhere and
one therefore expects to have a patchwork of regions with differing phases. If
on going round a loop in space which encloses several of these patches, the
phase changes by a non-zero multiple of $2 \pi$, then it follows that the
field must vanish at least once within the loop and therefore that there must
be at least one vortex passing through the loop. Thus defects are formed
between regions of differing phase.

The question of what determines the initial vortex density, however, is
less clear-cut.
How many defects are formed must depend on the length scale over which
the phase varies, or, rougly speaking, on the size of the domains of
approximately constant phase. Amongst cosmologists, it was previously thought
\cite{ginzburg}
that the relevant domain size was the thermal
equilibrium coherence length of the field at the Ginzburg temperature, $T_G$.
This is defined to be the temperature above which there is a significant
probability for a thermal fluctuation of the field on a coherence length
scale to unwind a defect by crossing the centre of the Mexican hat
potential. In other words:-
\begin{displaymath}
\xi^3 \Delta V / k T_G \approx 1,
\end{displaymath}
where $\Delta V$ is the difference in free energy density between the true and
false vacua.  The idea was that above
this temperature any strings would be unwound by thermal fluctuations
and only when the system cooled below this temperature could defects
become quasi-stable or `frozen in'.

This argument is now believed to be wrong. The phase transition and the
associated production of defects is an intrinsically non-equilibrium process,
so it is incorrect to compute the defect density using the equilibrium field
correlation length; indeed, it is strictly speaking not even possible
to talk of a well-defined temperature.  Even were it possible always to
define temperature, the Ginzburg
temperature would not be relevant \footnote{The problem of trying to describe
intrinsically non-equilibrium phenomena such as phase transitions, without
using equilibrium concepts such as an effective potential, is very generic.
For further comment see \cite{jackiw}.}.  The physical meaning of $T_G$ for the
vortices is that this is the temperature above which they become rough: the
thermal fluctuations in the vortex positions become larger than the vortex
width. It is not correct to argue, however, as has often been done, that above
this temperature strings can fluctuate in and out of existence.  Above $T_G$,
strings will start to wiggle randomly on small scales, and very small loops
may appear and disappear, but a long string is unlikely fluctuate in or out of
existence as this would require a fluctuation on a scale of several coherence
lengths. The small loops have very short lifetimes and would not
normally survive long enough to be seen. In other words it is only
coherent fluctuations on scales larger than the coherence scale of the
field which are long-lived.

What is really needed is a truly non-equilibrium approach to the calculation
of the defect density. In general, this is a difficult problem.  However, in
the case of a rapidly quenched second order phase transition, the Zurek
scenario, it is thought possible to make an estimate of the defect density
produced comparatively simply \cite{whz}. The current understanding of
this scenario is as follows.

In cosmological systems, there
is a bound on the possible size of phase-correlated domains due to
causality. A domain obviously cannot be larger than a horizon-sized
volume and if one assumes on dimensional grounds that the defect density
is roughly the reciprocal of the domain size squared, then this gives a
lower bound on the defect density produced. In condensed matter systems,
there is often an equivalent causal horizon which stems from the fact that
correlations in the order parameter are not only causally bounded in principle
by the speed of light, but in practice also by the speed at
which interactions propagate in the system concerned. This is the basis
of Zurek's estimate of the vortex density in the Lancaster experiment in
$^4$He and also of our work here.

Consider, for the sake of example, a condensed matter system like superfluid
$^4$He, which starts in thermal equilibrium above the critical temperature
and then undergoes a uniform temperature quench into the superfluid
symmetry-broken phase.  In real experiments a pressure quench is
usually involved as it is difficult to produce uniform temperature quenches.
For pedagogical clarity, however, we will consider first a temperature
quench and only later show how to deal with a pressure quench. As the
temperature is lowered and the phase
transition proceeds, initially  the relaxation rate of the field is greater
than the rate at which the  transition is being driven and the order
parameter is able to keep up with  the quench to maintain a roughly
equilibrium configuration. It is therefore possible to talk about the
temperature of the system during this early part of its evolution and also,
since the system is almost in thermal equilibium, to calculate the
correlation length of the field as it increases early in the transition.

As the second order transition is approached, however, for all modes of the
order parameter field there will come a point during the transition when the
relaxation of the order parameter is not fast enough to keep up with
the quench. This effect is particularly prominent during second order
transitions due to the critical slowing down of the low momentum modes
of the field. For sufficiently rapid quenches, all modes of the scalar field
will go out of equilibrium almost simultaneously and one can think of the
field configuration at this instant being frozen in until at some stage below
the transition the relaxation rate of the order parameter is again
greater than the rate at which the transition is being driven. During this
adiabatic phase of the transition, the dynamics will be roughly isentropic.
The domain structure for the phase of the scalar field which gives rise to
defects will then be that when the field configuration is frozen in
so the defect density is calculated by finding the correlation length
when the scalar field first goes out of equilibrium.
In reality of course, modes do not go out of equilibrium instantaneously.
Nevertheless, under the circumstances where the Zurek scenario is intended
to be applied, this model seems to give a reasonably accurate estimate
\cite{pvem}.

The Lancaster experiment in $^4$He discussed below uses a pressure- rather
than a temperature-quench to induce the phase transition. This situation is
experimentally simpler in that it avoids the difficulty of trying to induce a
uniform temperature quench.
We can assume that the process starts from an equilibrium state just above or
just below $T_c$, and there is a rapid pressure quench from there to a state
well away from the $\lambda$-transition at the lower pressure. In the spirit
of the Zurek scenario it then seems reasonable to assume that the equilibrium
configuration of the field at the original temperature and pressure is simply
frozen in when the field goes out of equilibrium and hence determines the
defect density measured.

\section{The Lancaster Experiment And Its Interpretation}

There have been three experiments relevant to the study of non-equilibrium
field theory in defect formation. The first involved a temperature
driven first order phase transition in nematic liquid crystals
\cite{neil}
and was primarily intended to
demonstrate the analogy between the behaviour of defect
networks in nematics long after the phase transition when the symmetry has
been broken almost everywhere and the scaling of
cosmic string networks. Although it
also had some relevance to the consideration of defect formation, it
suffered from the experimental problem that it is difficult to
produce a uniform temperature quench in a nematic over more than a few
correlation volumes.
The second experiment, performed in Lancaster \cite{pvem}, was
specifically designed to test the Zurek scenario and exploit the analogy
between a hypothetical grand unified scale
symmetry-breaking transition producing cosmic strings in the early
universe and the $\lambda$ transition in liquid $^4$He. Although this
second experiment is inherently far more difficult, preliminary results seem
to vindicate the Zurek scenario, subject to various caveats discussed below.
The third and most recent experiment \cite{mk},
involving superfluid $^3$He, also appears to vindicate the Zurek scenario.

\begin{figure}[htbp]
\epsffile{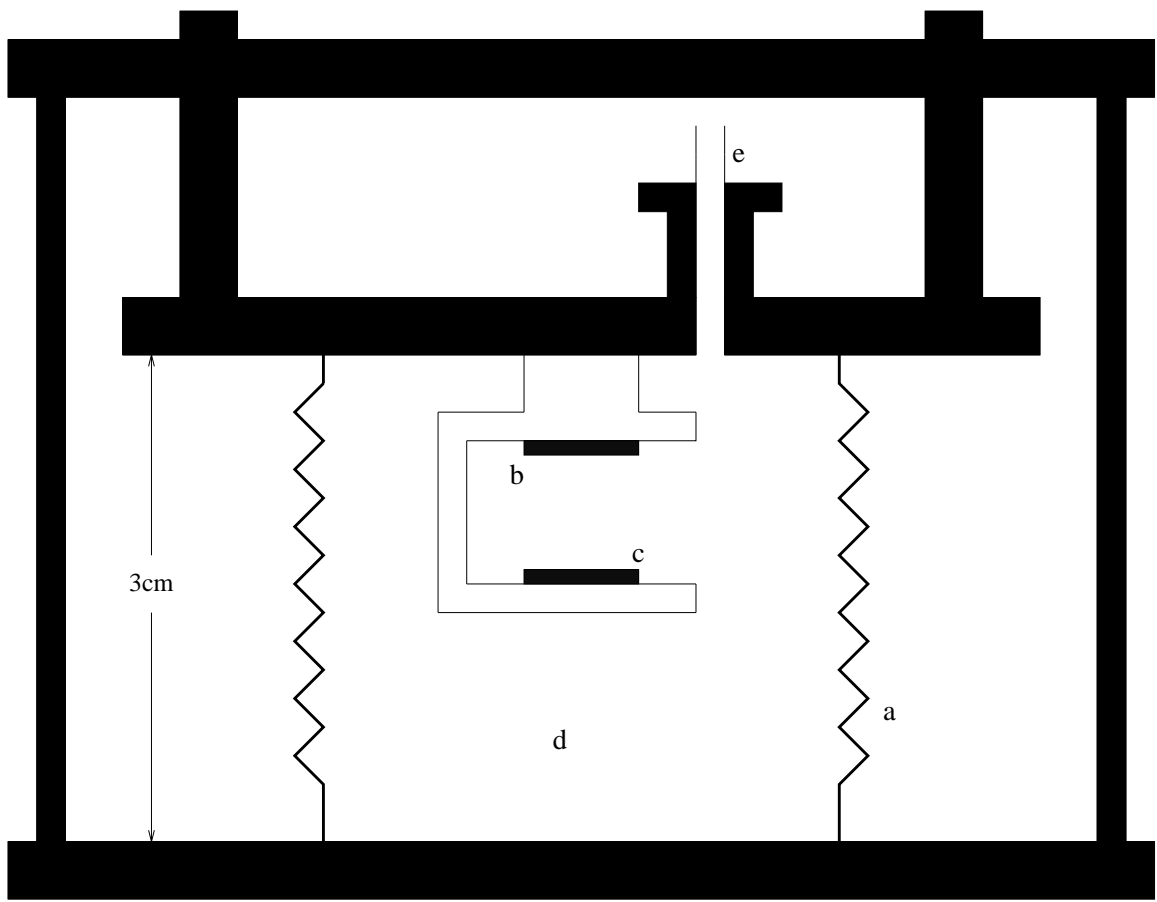}
\vskip 1cm
\caption{The Lancaster experiment (schematic --- not to scale):
$a$~phosphor-bronze bellows;
$b$~bolometer;
$c$~heater;
$d$~isotopically pure sample of $^4$He;
$e$~sample filling tube.}
\label{fig1}
\end{figure}

The Lancaster experiment which is shown schematically in Figure \ref{fig1}, may
be summarized as follows. A $10^{-3}$kg isotopically pure sample of liquid
$^4$He is held at about 2K in a bronze bellows a few centimetres in length.
Both the pressure and temperature of the sample are recorded and the vortex
density is measured indirectly by the attenuation of second sound between a
heater-bolometer pair four millimetres apart. The bellows is compressed and
the  sample allowed to reach thermal equilibrium before being released and
allowed  to expand by four millimetres in a period of roughly three
milliseconds. Since the speed of first sound, which carries information about
pressure and  therefore the progress of the transition, is so much greater
than that of second sound, which is related to the relaxation rate of the
scalar field, it is possible to induce the phase transition simultaneously
over many causally disconnected regions.

After the pressure quench, the vortex density produced will decay over a
period of a few seconds at a rate \cite{jv,schwarz,donnelly}
\begin{displaymath}
\frac{dL}{dt} = - \chi_2 \frac{\hbar}{m_{He}} L^2,
\end{displaymath}
where the Vinen parameter $\chi_2$ is a dimensionless constant and it has
been assumed that vortex creation due to the non-conservative interaction
between the normal fluid and vortices already present is not significant.
Although the whole apparatus vibrates for a few tens of milliseconds after
the quench thus initially swamping the electronic detection apparatus,
and the decay of such a vortex density is rapid, using the
defect attenuation of second sound, which travels at a few metres a
second in the regime concerned, there is time to measure the
decreasing vortex density before equilibrium is reached and the attenuation
becomes constant.

There are several problems with this experiment, most obviously the
effects of turbulence introduced by the walls of the bellows and the capillary
used to fill the cell, the latter possibly being the more serious effect.
Also, it is difficult to repeat exactly the same run many times as the precise
isentrope followed during the quench can only be determined retrospectively.
Although a pressure quench induced by allowing the bellows to expand
produces a very uniform quench, the vibrations of the bellows mean that
the electronics of the heater-bolometer system are swamped by noise for
about forty milliseconds.
It has also been commented that it is very difficult to prepare a sample
of superfluid helium-4 without any vorticity and that it is therefore
not entirely clear what is causing the vorticity.

Regardless of the problems, however, the results are as follows.
For quenches from above the lambda transition, the expected
attenuation of the second sound signal is seen. Although at very early times
after the transition, the electronics are swamped by spurious signals due to
the vibration of the apparatus, it is straightforward to extrapolate although
unfortunately only to a limit on the defect density formed of
\begin{displaymath}
n > 10^{13} {\rm m}^{-2},
\end{displaymath}
where $n$ is the vortex line density. The Zurek prediction is close to this
limit.  These results do allow us at least to determine that the Ginzburg
temperature is not relevant. The Ginzburg temperature lies far below the
final state to which the system is quenched, about half a degree Kelvin
below the critical temperature and can not therefore be
relevant to the formation of defects. The Zurek prediction is seemingly very
good, but the presence of vorticity after quenches from just below the
$\lambda$-transition casts doubt on the Zurek scenario as the only
significant means of vortex production.

Similar quenches starting from thermal equilibrium significantly below the
$\lambda$-transition, but still within the
Ginzburg regime, do not show an equivalent
attenuation and it is claimed that this rules out vortex production by
motion between the fluid and the corrugated walls of the bellows \cite{pvem}.
If, however, the expansion which induces the pressure quench commences
within about 10mK of the transition, some comparatively short-lived
attenuation {\it is seen} and indicates the presence of vortices although at
a lower density than that produced by quenches through the $\lambda$-
transition. It is this vorticity which concerns us here.

\section{Quenches From Below $T_{\lambda}$}

Clearly the defects which are observed after quenches starting from below the
phase-transition are not formed by the Kibble mechanism since the symmetry
is already broken in the initial state. This forces us
to question whether the vorticity created during quenches from above the
phase transition is due to the Kibble mechanism or whether there is some
additional means of forming defects which contributes to the vorticity
created in transitions from both above and below the phase-transition.
The most obvious idea is to attribute the sub-critical vortices to thermal
fluctuations in the order parameter. In other words, so close to the phase
transition the thermal defect density is sufficiently high that these defects
live through the quench to be seen in the Lancaster experiment. The fact
that such vortices would be almost all in the form of small coherence sized
loops is irrelevant as they will all still attenuate second sound. Let
us therefore estimate this thermal defect density. The first guess would
therefore be that, although with a quench from a point already below the phase
transition critical slowing down becomes less significant as the transition
proceeds, rather than more important as was the case with the original Zurek
scenario, for
sufficiently fast quenches, the thermal vortex density is frozen in and ends
up being the observed density. For more realistic quenches at slower rates,
the system will be better able to keep up with the quench and will therefore
end up closer to the perfect equilibrium situation with fewer vortices than
would be predicted by this freezing- in argument. In other words, applying
the Zurek scenario to quenches either from above or below the transition
gives an upper limit on the vortex density.

In order to compute the density of vortices, we need some way to
identify and count configurations of the field which have non-zero winding
number and constitute vortices. If on
going round a loop in space, the phase of the scalar field changes by
some non-zero integral multiple of $2 \pi$, then it is not possible for
the field to be in the true vacuum state everywhere within the loop.
There must be at least one point within the loop where the field is in
the false vacuum, that is where the field vanishes. This point can be
thought of as the centre of the string and is the only gauge invariant
value of the field. Thus, in any gauge, the centre of a string must be a
zero of the field. Although it is true that all strings have a zero at their
centre, it is not necessarily true that all zeroes of the field are associated
with vortices. What is in fact required is the number density of zeroes
with large-scale winding around them.

Halperin has suggested a method of computing the topological line density
$\rho(\bf r)$ or defect density \cite{halperin,mazenko} defined by
\begin{displaymath}
\rho({\bf r}) = \sum_{n}\int ds \frac{d{\bf R}_{n}}{ds}
\delta^{3} [{\bf r} - {\bf
R}_{n}(s)],
\end{displaymath}
where $ds$ is the incremental length along the line
of zeroes ${\bf R}_{n}(s)$ ($n$=1,2,.. .) and $d{\bf R}_{n}/{ds}$ is a unit
vector pointing in the direction which corresponds to positive
winding number.  Only winding numbers $n = \pm 1$ are considered.
Higher winding numbers are understood as describing multiple zeroes.
If $dA$ is an incremental two-dimensional surface containing the
point ${\bf r}$, whose normal is in the $j$th direction, then
$\rho_{j}({\bf r})$ is the net density of strings or in other words the
density of strings {\it minus} the density of antistrings on $dA$.

We shall only consider situations in which
\begin{displaymath}
\langle\rho_{j}({\bf r})\rangle = 0.
\end{displaymath}
That is, we assume equal likelihood of a string or an antistring passing
through an infinitesimal area.
It follows that, in terms of the zeros of $\Phi ({\bf r})=\Phi_1({\bf r})
+ i\Phi_2({\bf r})$, $\rho_{i}({\bf r})$ can be written as:-
\begin{displaymath} \rho_{i}({\bf r}) = \delta^{2} [\Phi ({\bf r})]
\epsilon_{ijk} \partial_{j} \Phi_{1}({\bf r}) \partial_{k}\Phi_{2}({\bf r}),
\end{displaymath}
where
$\delta^{2}[\Phi ({\bf r})] = \delta[\Phi_{1} ({\bf r})] \delta[\Phi_{2}
({\bf r})]$.
The coefficient of the $\delta$-function in this expression is
the Jacobian of the transformation from line zeroes to field zeroes.
While the above expression is very good insofar as it counts only
objects with winding which are genuinely defects, it is not always the
easiest thing to calculate.
In thermal equilibrium, however, the gradient terms take on an average value
and Halperin's method is then entirely equivalent to counting the thermal
expectation of the number of zeroes as follows.

In our simplified method, which applies only to thermal equilibrium when the
structure of the vortices is known, we count all the zeroes of a smoothed
field on a particular
surface, regardless of sign.  In thermal equilibrium at a temperature $T$
below the phase transition, the expected number density  of defects passing
through a circular loop of radius $R$ will be:-
\begin{displaymath}
\langle n \rangle = \frac{1}{\pi R^2}
\biggl \langle \, {\rm number\ of\ zeros\ of}\
\phi \, \biggr \rangle,
\end{displaymath}
where the triangular brackets denote a quantum mechanical and thermal
average which must ultimately be computed using some field theory
describing $^4$He.

One advantage of considering the zeros of $|\phi|^2$ is that this is a
manifestly gauge-invariant quantity. To count the zeros, we look for
regions where $|\phi|^2$ is within some specified range of zero,
namely $|\phi| \leq \eta(T)/ \sigma$, where $\eta(T)$ is the vacuum
expectation value of the field in the broken phase --- we may call this
the region of hot phase. Ultimately we take the limit $\sigma\to\infty$. For
small values of $r$, an individual vortex is decribed by
\begin{displaymath}
|\phi(r)| \approx \eta(T)r / \xi(T),
\end{displaymath}
where $1 / \xi(T)$ is the temperature-dependent mass of the scalar
particle which can later be calculated from the action \cite{wiegel}.  Hence,
assuming an isotropic distribution of the thermal string and averaging over
the possible orientations of such string, which give different areas of
intersection with the disc of radius $R$, the area within which the hot-phase
condition is satisfied around a single vortex is
\begin{displaymath}
A_{\rm string} = \frac {2 \pi \xi^2} {\sigma ^2},
\end{displaymath}
where the factor of two comes from the angular averaging process. It follows
that \begin{displaymath}
\langle n \rangle = \frac{1}{\pi R^2}
\frac {\langle {\rm Area\ in\ hot\ phase} \rangle }
{A_{\rm string} }.
\end{displaymath}

To find the area in the hot phase, we use a Gaussian window function
$\exp[-\sigma^2|\phi|^2/\eta^2(T)]$.  Thus the expected number density of
zeros is
\begin{eqnarray}
\nonumber
\langle n \rangle &=&
\frac{1}{\pi R^2} \lim_{\sigma \rightarrow \infty} \frac{\sigma
^2}{\pi \xi^2(T)} \Biggl \langle \int_{\rm loop} d^2{\bf x} \exp
\biggl ( \frac{-
\sigma^2|\phi|^2}{\eta^2(T)}\biggr )
\Biggr \rangle
\\
\nonumber
&=& \frac {1}{\pi \xi^2(T)}
\lim_{\sigma \rightarrow \infty} \Biggl
\langle \sigma^2\exp \biggl (\frac{-\sigma^2
|\phi({\bf 0})|^2} {\eta^2(T)}
\biggr ) \Biggr \rangle,
\label{nexp}
\end{eqnarray}
since $\langle |\phi|^2 \rangle$ is translationally invariant.

Actually, we are not interested in all zeros.  Small-scale
fluctuations may lead to the appearance of several short-lived zeros
within a particular vortex.  To eliminate over-counting, we will
smooth $\phi$ on a scale comparable to the thermal coherence length.
Indeed, the assumed behaviour of the field near a zero is correct only
if the field is smoothed. Thus we avoid counting two zeros within a string
width of each other as more than one string. We do this by replacing
$|\phi({\bf 0})|^2$ with
\begin{equation}
\int_0^\beta d \tau \int d^3{\bf x}
\int_0^\beta d \tau' \int d^3{\bf y} \,\,\,
\phi^*({\bf x}) f({\bf x}) \delta(\tau)  \phi({\bf y}) f({\bf y})
\delta(\tau'),
\label{phisq}
\end{equation}
where
\begin{displaymath}
f({\bf x})=
\Biggl ( \frac{1} {2 \pi \zeta^2} \Biggr )^{3/2}
\exp \Biggl ( \frac { -|{\bf x}|^2 } {2 \zeta^2 } \Biggr ).
\end{displaymath}
Here $\zeta$ is a the length-scale on which the field is smoothed. It is
necessary to have a very good physical motivation for the choice of this
scale. Later, it will be chosen to be of the order of the thermal correlation
length $\xi(T)$. It need not a priori be equal to it, however.
Since $f({\bf x})$ is strongly peaked within a thermal
coherence length of the origin but still non-singular, this technique has the
additional advantage of eliminating unphysical divergences in the thermal and
quantum mechanical averages. On scales smaller than a coherence length, the
quantum fluctuations will dominate over the thermal. In other words, on
such scales the Heisenberg uncertainty principle is producing a large
population of virtual defects which would cause a divergence in the
number density without a cut-off. This is very similar to a population
of virtual photons round an electron, although it is more obvious if
one uses a dual theory of the transition in which the quanta are the vortices
instead of the usual field theory where the quanta are
related to helium atoms.

The number density of
vortices can therefore be found by calculating the expected number of
zeros of the scalar field on a hypothetical flat circular disc. This is close
to
the
experimental procedure, where the defect density is measured by the
attenuation of second sound, which is attenuated by any region of
symmetry-unbroken phase. It should be noted that this method will not in
itself give any information about the length distribution of the vortices.

Assuming that the smoothed scalar field does not vary very much over
length scales comparable with the range of the pair potential $V({\bf x})$,
which experimentally is of the order of the typical atomic
separation, we are able to use an action with the local form
\begin{equation}
S[\phi] = \int_0^{\beta} d \tau \int d^3{\bf x} \, \,
\phi^* ({\bf x}) \biggl [ - \frac{1}{2m} \nabla^2 - \mu
+ \frac{\partial}{\partial \tau} + \lambda
\, \, |\phi({\bf x}) |^2 \biggr ] \phi({\bf x}),
\label{action}
\end{equation}
where $\lambda = \tilde V({\bf 0})$ and we set $\hbar=1$. SI units will
be restored later in order to facilitate a comparison with experiment.
Also, the term which looks like a chemical potential, $\mu$, is defined
so that $\mu = \mu_0 + V(0)/2$ where $\mu_0$ is the genuine chemical
potential of the helium atoms.
This may be deduced from many particle quantum mechanics with the sole
additional approximation that the interactions may be represented
locally. This may not be a particularly good approximation, however, since
the helium
atoms have a finite  size and separation of the order of a few \AA ngstroms
not much less than a typical coherence length of the field, thus placing a
limit on how local the  interaction can be. Other
condensed matter systems, such as superfluid $^3$He, do not suffer from
this problem. They do, however, exhibit a variety of defects which makes
calculation significantly more complicated if no more difficult in principle.

\section{Calculation Of The Thermal Density}

Consider first the calculation of the vortex density in thermal
equilibrium in the symmetry-broken phase. Below the temperatures at which the
phase transition actually occurs, we may expand the field $\phi$ about a
particular point on its vacuum manifold
\begin{displaymath}
\sqrt{2} \phi = \eta(T) + A({\bf x}) + i B({\bf x}).
\end{displaymath}
Let us for the moment assume that we may use mean field theory, and so
rewrite the combinations of $A$ and $B$ thus:
\begin{eqnarray}
\nonumber
A^3 &\approx& 3 \langle A^2 \rangle A,
\\
\nonumber
A^4 &\approx& 6 \langle A^2 \rangle A^2
- 3 \langle A^2 \rangle ^2,
\\
\nonumber
A^2 B^2 &\approx& \langle A^2 \rangle B^2
+ A^2 \langle B^2 \rangle -
\langle A^2 \rangle \langle B^2 \rangle,
\\
\nonumber
AB^2 &\approx& A \langle B^2 \rangle,
\end{eqnarray}
where quantities such as $\langle A^2 \rangle$ are later to be computed
self-consistently. It is quite reasonable to object to this in that the
region of interest is well within the Ginzburg regime in $^4$He, in other
words less than half a degree Kelvin away from the transition. This region
is defined to be precisely the region in which thermal fluctuations are
important and mean field theory is known to be a poor approximation in this
case \cite{ng}.
However, mean field theory will give at least
qualitatively the right behaviour and is unlikely to be inaccurate by
{\it many} orders of magnitude. It turns out that this is sufficient for our
purposes.

We may now use a path integral method to compute the
number density of zeros.  The integrand of the path integral contains the
exponential factor from (\ref{nexp}), with $|\phi({\bf 0})|^2$ replaced by
(\ref{phisq}), as well as the exponential of the action (\ref{action}).  The
integral factorizes, in the form
\begin{equation}
\int {\cal D} A  \exp \biggl \{ -  S[A] \biggr \}
\int {\cal D} B  \exp \biggl \{ -
S[B] \biggr \} =  \,I_A \, I_B,
\label{iaib}
\end{equation}
where
\begin{eqnarray}
\nonumber
S[A] &=& \int_0^{\beta} d \tau \int d^3{\bf x} \int_0^{\beta} d
\tau' \int d^3{\bf y} \,
\Biggl [
  \frac{\sigma^2}{2 \eta^2}
    \biggl (
      \eta^2 + A({\bf x}) A({\bf y}) + 2 \eta A({\bf y})
    \biggr )
  f({\bf x}) f({\bf y}) \delta(\tau) \delta(\tau')
\\
\nonumber
  &+& \frac{ A({\bf x}) } {2}
  \biggl (
    -\frac{1}{2m} \nabla^2
     +  \frac{\partial}{\partial \tau} + C_2
  \biggr )
  \delta^3 ({\bf x} - {\bf y}) \delta(\tau - \tau') A({\bf y})
\Biggr ]
- \int_0^{\beta} d \tau \int d^3{\bf x}
\, C_1 A({\bf x})
\end{eqnarray}
and
\begin{eqnarray}
\nonumber
S[B] &=& \int_0^{\beta} d \tau \int d^3{\bf x} \int_0^{\beta} d
\tau' \int d^3{\bf y} \,
\Biggl [
  \frac{\sigma^2}{2 \eta^2}
    \biggl (
      B({\bf x}) B({\bf y})
    \biggr )
  f({\bf x}) f({\bf y}) \delta(\tau) \delta(\tau')
\\
\nonumber
  &+& \frac{ B({\bf x}) } {2}
  \biggl (
    -\frac{1}{2m} \nabla^2
    +  \frac{\partial}{\partial \tau} + C_3
  \biggr )
  \delta^3 ({\bf x} - {\bf y}) \delta(\tau - \tau') B({\bf y})
\Biggr ],
\end{eqnarray}
where the constants, $C_1$, $C_2$ and $C_3$ are defined as
\begin{eqnarray}
\nonumber
C_1 &=& \lambda \eta \, [ \, \eta^2
+ 3 \langle A^2 \rangle + \langle
B^2 \rangle \, ] - \mu \eta,
\\
\nonumber
C_2 &=& \frac{\lambda}{2} \, [ \, 3 \eta^2
+ 3 \langle A^2 \rangle + \langle
B^2 \rangle \, ] - \frac{\mu}{2},
\\
\nonumber
C_3 &=& \frac{\lambda}{2} \,
[ \, \eta^2 + \langle A^2 \rangle + 3 \langle
B^2 \rangle \, ] - \frac{\mu}{2}.
\end{eqnarray}
These path integrals are of standard well-defined Gaussian
type and since the non-perturbative physics is already in our equations,
it does not matter that we expand about a particular
point on the vacuum manifold.  Before proceeding to do
the path integrals, however, let us first consider the self-consistent
calculation of the coefficients $C_1$, $C_2$ and $C_3$, or equivalently of
$\langle A^2 \rangle$ and $\langle B^2 \rangle$. To do this, we exploit the
fact that the real scalar fields $A$ and $B$ must have a null expectation
value. Viewing $C_1 A$ as a term in the potential for the $A$ field, it will
be seen that the only way to achieve $\langle A \rangle = 0$
self-consistently is to impose $C_1 \equiv 0$. This then enables us to
calculate $C_2 = \lambda \eta^2$.

We also know that $B$ is a Goldstone mode, which implies
that $C_3 = 0$. An interesting corollary of the fact that $C_1$ and
$C_3$ both vanish is that $\langle A^2 \rangle = \langle B^2 \rangle$.
Physically, this means that when both modes have zero effective
mass at the lambda point, the equipartition theorem forces them
to have the same average energy.


The result of this determination is to greatly simplify the $A$ and
$B$ integrals. The path integrals may now be evaluated directly as they are
both Gaussian. We will normalize both by
dividing by similar integrals with the parameter $\sigma$ set to zero,
thus obtaining the actual number density of vortices. Quantities in
which $\sigma = 0$ will be denoted by a zero subscript.

Consider first the $B$ integral. This is of the form
\begin{equation}
I_B = \int {\cal D} B \, \exp \biggl
( - \frac{1}{2} B K B \biggr )
=(\det K)^{-1/2},
\label{ib}
\end{equation}
where we have adopted the integration convention that spatial and
Euclidean time integrals are now implicit on adjacent fields and
operators and
\begin{displaymath}
K({\bf x},\tau;{\bf y},\tau') = \frac{\sigma^2}{\eta^2} f({\bf x}) f({\bf y})
\delta (\tau) \delta(\tau') + \biggl (
-\frac{1}{2m} \nabla^2  +  \frac{\partial}{\partial \tau}
\biggr ) \delta^3 ({\bf x} - {\bf y}) \delta(\tau - \tau').
\end{displaymath}
The Fourier transform of this,
\begin{displaymath}
\tilde{K}(p,q) = \frac{\sigma^2}{\eta^2} \tilde{f}({\bf
p}) \tilde{f} ({\bf q}) + \biggl (
\frac{|{\bf p}|^2}{2m} + i p_0
\biggr ) (2\pi)^3\delta^3({\bf p} - {\bf q})\, \beta\delta_{p^0,q^0}.
\end{displaymath}
allows us to compute the ratio $I_B/I_{B,0}$.  This gives
 \begin{eqnarray}
\nonumber
\Biggl ( \frac{I_B}{I_{B,0}} \Biggr )^{-2}
&=& \det \biggl (\frac{K}{K_0} \biggr )
\\
\nonumber
&=& \exp \sum_{n=1}^\infty\frac{(-1)^n}{n} {\rm Tr}
\biggl [  K_0^{-1} ( K-K_0 ) \biggr
]^n
\\
\nonumber
&=& \exp \sum_{n=1}^\infty\frac{(-1)^n}{n} \biggl(
{\sigma^2\over\eta^2}\int{d^3{\bf p}\over(2\pi)^3}
{1\over\beta}\sum_{r=-\infty}^{\infty}
\frac{|\tilde{f}({\bf p})|^2
}{|{\bf p}|^2 / 2m + 2\pi ir/\beta}\biggr)^n
\\
\nonumber
&=& 1+ \frac{\sigma^2}{\beta \eta^2}
\sum_{r=-\infty}^\infty\int {d^3{\bf p}\over(2\pi)^3}
 \frac { | \tilde{f} ({\bf p}) |^2 }
{ | {\bf p} |^2 / 2m + 2 \pi ir / \beta }.
\end{eqnarray}
Using the known result
\begin{displaymath}
\frac{2 x}{\pi} \sum_{n=1}^\infty
\frac{1}{x^2 + n^2} = \coth ( \pi x)
- \frac{1}{\pi x},
\end{displaymath}
we may rewrite the sum to give
\begin{displaymath}
\frac{I_B}{I_{B,0}} = \biggl (
1 + \frac{\sigma^2D_0}{\eta^2}\biggr)^{-1/2},
\end{displaymath}
where
\begin{displaymath}
D_0={1\over2}\int {d^3{\bf p}\over(2\pi)^3}
|\tilde{f}({\bf p})|^2 \coth \biggl
( \frac {|{\bf p}|^2} {4mT} \biggr ),
\end{displaymath}
with
\begin{displaymath}
\tilde{f}({\bf p}) =
\exp \biggl ( \frac{- \zeta^2 |{\bf
p}|^2 } {2} \biggr ).
\end{displaymath}
If $\zeta^2\gg1/mT$, we obtain
\begin{displaymath}
\frac{I_B}{I_{B,0}} \approx \biggl(
1 + \frac{\sigma^2 m T}{2\pi^{3/2}\eta^2\zeta}
\biggr)^{-1/2}.
\end{displaymath}

The corresponding integral for the $A$ field is slightly more
complicated and of the form
\begin{eqnarray}
\nonumber
I_A &=& \int {\cal D} A \exp \biggl \{
- \frac{1}{2} AKA - bA - {\sigma^2\over2}\biggr \}
\\
&=&\biggl [ \det K \biggr ]^{-1/2} \,
\exp \biggl ( \frac{1}{2} b K^{-1}
b -{\sigma^2\over2}\biggr ),
\label{ia}
\end{eqnarray}
where again there is implicit integration on adjacent fields and
operators and
\begin{eqnarray}
\nonumber
K({\bf x},\tau;{\bf y},\tau') &=& \frac{\sigma^2}
{\eta^2} f({\bf x}) f({\bf y})
\delta (\tau) \delta(\tau') + \biggl (
-\frac{1}{2m} \nabla^2
+  \frac{\partial}{\partial \tau} + \lambda\eta^2
\biggr ) \delta^3 ({\bf x} - {\bf y}) \delta(\tau - \tau'),
\\
\nonumber
b({\bf x},\tau) &=& \frac{\sigma^2 f({\bf x}) \delta(\tau)}{\eta}.
\end{eqnarray}
The required result is
\begin{displaymath}
\frac{I_A}{I_{A,0}} = \Biggl [ \det
\biggl ( \frac{K}{K_0} \biggr
)\Biggr ]^{-1/2} \exp \biggl (
\frac{1}{2}  b K^{-1} b - {\sigma^2\over2}\biggr ).
\end{displaymath}
The determinental factor follows as for the $B$ integral:
\begin{displaymath}
\det \biggl ( \frac{K}{K_0} \biggr)
 = 1 + \frac{\sigma^2}
{2 \eta^2} \int {d^3{\bf p}\over(2\pi)^3}
\, | \tilde{f}({\bf p})|^2 \coth
\biggl [ \frac{|{\bf p}|^2}{4mT} +
\frac{\lambda\eta^2}{2T} \biggr ].
\end{displaymath}
The exponential factor involves more extensive manipulations:
\begin{eqnarray}
\nonumber
bK^{-1}b  &=& \frac
{ \sigma^2 f} {\eta}
\, \Biggl (
K_0^{-1} - \frac{ \sigma^2
f K_0^{-1} f } {\eta^2 + \sigma^2
f K_0^{-1} f } K_0^{-1}
\Biggr ) \,
\frac { \sigma^2 f} {\eta}
\\
\nonumber
&=& \frac {\sigma^4 D} {\eta^2 + \sigma^2 D},
\end{eqnarray}
where
\begin{displaymath}
D = f K_0^{-1}f = \frac{1}{2}\int {d^3{\bf p}\over(2\pi)^3}
\, | \tilde{f}({\bf p})|^2 \coth
\biggl [ \frac{|{\bf p}|^2}{4mT} +
\frac{\lambda\eta^2}{2T} \biggr ].
\end{displaymath}
Substituting this into the expression for the normalized $A$ integral
above yields
\begin{displaymath}
\frac{I_A}{I_{A,0}} = \biggl
(1+\frac{\sigma^2D}{\eta^2} \biggr)^{-1/2}
\exp \Biggl [
\frac{ - \eta^2 \sigma^2 } { 2( \eta^2 +
\sigma^2 D ) } \Biggr ].
\end{displaymath}

Combining both of the factors (\ref{ia}) and (\ref{ib}) into (\ref{iaib})
gives the final expression for the number density of vortices passing through
a unit area at temperature $T$ as
\begin{displaymath}
\langle n \rangle=\lim_{\sigma
\rightarrow \infty} \frac{\sigma^2}{2 \pi\xi^2(T)}
\biggl( 1 + \frac{\sigma^2D_0}{\eta^2} \biggr)^{-1/2}
\biggl( 1 + \frac{\sigma^2D}{\eta^2} \biggr)^{-1/2} \,
\exp \Biggl [
\frac{ - \eta^2 \sigma^2 } { 2( \eta^2 + \sigma^2 D ) }
\Biggr ].
\end{displaymath}
On taking the limit, we obtain the ultimate result
\begin{equation}
\langle n \rangle = \frac{\eta^2(T)}
{2 \pi\xi^2(T)(D_0D)^{1/2}} \exp \Biggl
( \frac{-\eta^2(T)}{D} \Biggr ).
\label{n}
\end{equation}
The integrals in $D$ and $D_0$ do not have any closed analytic form but
are trivial numerically. However, if the temperature is
large in the sense that $T \gg 1 / 2m \xi^2$ or in other words, $T \gg \mu
= \lambda \eta^2$, which is true well within the region of interest, the
Ginzburg regime, then (\ref{n}) reduces to the approximate form
\begin{equation}
\langle n \rangle = \frac{ \eta^2(T) \zeta}
{\pi\xi^2(T)mT} \exp \Biggl
( \frac{-\eta^2(T)\zeta}{mT} \Biggr ).
\label{napp}
\end{equation}

This result depends strongly on the smoothing scale $\zeta$. Conventionally in
quantum field theory, one expects first to regulate an apparently divergent
result by introducting a cut-off scale and then to show that the physically
observable quantity is not dependent on this scale by renormalising. In the
current context, however, the physically relevant quantity should depend on a
cut-off for two reasons. Firstly there is an intrinsic cut-off in the problem
due to the discrete nature of the superfluid. Clearly, it is not sensible to
talk
about superfluid flow forming a vortex on scales smaller than this. In
addition,
however, we are only interested in coherent flows of the superfluid which are
sufficiently long-lived to be seen by, for example, scattering experiments like
that involving second sound described above. In other words we are only
interested in flows of the superfluid on scales larger than the coherence
length.
There will indeed be fluctuations on scales smaller than this, which have
winding
number and zeroes, but they are not relevant to the Lancaster experiment as
this
sees vortices only on longer time-scales. This is easiest to understand in
terms of a classical analogy with a hurricane. Like a quantum vortex in
superfluid $^4$He, a hurricane is a large-scale coherent motion of a fluid. One
can clearly not talk about vortex or hurricane-like motions on scales less than
the size of the particles making up the air. There will, however, be small
scale eddies on scales less than the large-scale coherent rotation of the whole
hurricane. When counting hurricanes, however, it would not be considered even
vaguely sensible to include the small-scale, short-lived eddies within the
main hurricane. For this reason, we choose our smoothing scale $\zeta$ to be
the thermal coherence length in our theory, $\xi(T)$. On restoring the factors
of
the Boltzmann and Planck constants, this leaves the following
expression for the thermal vortex density:-
\begin{equation}
\langle n \rangle = \frac {1}{\pi \xi^2 } \frac{ \eta^2(T) \xi(T) }
{k_B mT} \exp \Biggl
( \frac{ -\eta^2(T) \xi(T) }{k_B mT} \Biggr ).
\label{coh}
\end{equation}
This is not surprising:
the energy of a correlation-sized loop of string is of order $\eta^2\xi/m$,
so the exponent here is essentially the expected Boltzmann factor.

However, the above calculation of the defect density depended crucially on
the use of mean field theory to decouple the Goldstone and Higgs modes of
the scalar field and to force the integrals into Gaussian form.
Although this is perfectly valid well below the transition, outside the
Ginzburg regime, as we have already noted, it is not applicable to the main
region of interest near the $\lambda$ point. This is the major problem with
our method of calculation. It will turn out, however, that even with the most
optimistic assumptions concerning the superfluid density and coherence length,
that the thermal defect density is too small by at least a few and possibly by
many orders of magnitude to be the sole explanation for the vorticity observed
after sub-critical quenches. Mean field theory is therefore sufficient for our
purpose.

As a somewhat academic point we could in principle try to use the
renormalisation group to extrapolate into the region of interest. In the
current context, however, it is not entirely clear how
to go about applying these techniques. Physically, one expects that the form of
the result (\ref{napp}) is unlikely to change discontinuously on reaching the
Ginzburg regime and is likely always to be of the form:-
\begin{displaymath}
\langle n \rangle = \frac{\eta^2(T) }
{\pi\xi^2(T)} \exp \biggl (
\frac{-\eta^2(T)}{\langle A^2 \rangle_f}
\biggr ) \, \langle A^2
\rangle_f^{-1/2} \, \langle B^2 \rangle_f^{-1/2},
\end{displaymath}
but with different expression for ${\langle A^2 \rangle_f} = D$ and
${\langle B^2 \rangle_f} = D_0$. One might therefore try to merely replace the
coherence length and order parameter with renormalisation group improved
values.

In this context, a very interesting point emerges from (\ref{napp}) if we
set $\zeta=\xi(T)$ and consider the limit as $T\to T_c$.  In mean field theory,
$\eta^2\xi\propto \epsilon^{1/2}$, where $\epsilon$ is the reduced temperature
$\epsilon = 1 - T/T_c$, so the number
of vortices per correlation area would tend rather rapidly to zero as $T\to
T_c$.  On the other hand, in the renormalization group, $\eta^2\xi \propto
\epsilon^{2\beta-\nu}$.  But $2\beta$ is actually very nearly  equal to $\nu$
(they would be exactly equal if the correlation-function critical exponent
$\zeta$ were zero, or equivalently if $\delta=5$, in the usual notation). The
best estimate is $2\beta-\nu\approx 0.03$.  This means that until we get
extremely close to the critical point, the number of vortices per correlation
volume would remain of order one.

It is, however, difficult to decide what is the most honest, consistent thing
to do. Let us make the obvious assumption that the vortex density seen
immediately after sub-critical quenches is just the frozen in thermal
density. For quenches from within a few mK of the transition, starting at a
relative temperature of $\epsilon = 1- T/T_c = 10^{-3}$, the thermal vortex
line density calculated in the above mean field scheme is of order
$\langle n \rangle \approx 10^5 {\rm m}^{-2}$. Even closer to the transition
at $\epsilon = 10^{-4}$ the density is still only about
$10^{10} {\rm m}^{-2}$. While this is consistent in the sense that mean field
theory has been used throughout, there is some justification for trying to
use renormalisation group improved values.

The other obvious thing to try is to simply substitute experimental values for
the required quantities \cite{ahlers,wheatley,donnelly}, using the molar volume
as a function of pressure and the superfluid fraction to extract the vacuum
expectation value of the order parameter and tabulated values for the coherence
length. Even with optimistic values for these parameters, the thermal defect
density in the region of interest is still only of the order of
$10^{10}$m$^{-2}$ at best.

One's first thought is that perhaps the mean field approximation is so bad in
the regime of interest, well above the Ginzburg temperature, as to
severely underestimate the thermal defect density. Although the mean
field approximation is certainly not good in the regime of interest, it
is hard to believe that it is in error by orders of
magnitude, so our estimation of the thermal density seems to rule out the
the thermal vortex population as the sole source of vorticity from sub-lambda
quenches. There are two possibilities, although these are not mutually
exclusive. Either the thermal density acts as a seed and is
magnified in some way or else there is some other source of vorticity.

\section{Alternative Explanations Of The Sub-Critical Vortex Density}

Let us first consider other sources of vorticity. The most likely source is
the flows which are generated during the pressure quench by either the walls
of the bellows or the capillary used to fill the bellows in the first
place. The phosphor-bronze bellows are
corrugated in order to allow for the compression and subesequent release of
the sample chamber which produces the pressure quench. The capillary used to
fill the sample chamber has to be long to ensure thermal isolation and has a
valve on the end furthest from the opening into the sample. Thus during
the pressure quench the column of fluid inside the capillary expands and
about 20\% of its volume is forcibly squirted into the chamber \cite{pvem2}.

An idea due to Zurek is as follows \cite{whz2}.
The difference in the line density of vortices produced in quenches from very
close to the transition and the density formed from slightly further away
from the transition might be due to the different energy cost of creating
vortices. It takes less energy to
create a unit length of string near the transition since the vortex
energy per unit length decreases with the superfluid density. The nearer the
$\lambda$-transition the system is, the lower the superfluid density and
hence the lower the tension of a vortex and the easier it is to form a
unit length of one. The line density of vortex created in this way would
then be proportional to the reciprocal of the superfluid density, or
including logarithmic terms \cite{whz2}
\begin{displaymath}
\langle  \rho_s n \rangle_{ {\rm stir} } \propto
\biggl (
- \ln ( \langle n \rangle_{ {\rm stir} } \xi^2 )
\biggr )^{-1}.
\end{displaymath}
This ought to be testable by comparing sub-critical quenches with
differing starting points. Since this argument does not depend in any way on
the microphysics of superfluids, it is very robust but consequently it also
fails to provide any insight into the details of what is actually going
on and makes it difficult to ascertain whether the Zurek scenario is still
valid in this regime.

Given the presence of such flows during the pressure quench, however, one
might speculate that they interact with the thermal density of vortices
which is already present. The phenomenology of this is well understood.
Also, the interaction of a superflow with thermally generated vortices is
well known in the context of the thermal generation of turbulence in a
superfluid and is known by the acronym `ILF' after Iordanskii, Langer and
Fisher \cite{donnelly2}. The presence of a filling
capillary in the middle of the bellows is guaranteed to generate a superflow
during the rapid pressure decrease of the quench and this capillary is
probably far worse than the walls in terms of flow generation. Although the
walls in principle allow vorticity to be generated by their motion relative
to the superfluid, the rapid decompression of the superfluid in the capillary
during the quench causes a superflow from the capillary.

Let us assume that our estimate of the thermal density of vortices
just below the transition is reasonably accurate, at least to within a
factor of ten or so. Since they are a thermal equilibrium population, these
vortices will almost all be in the form of small coherence length sized
loops and will move through the superfluid with a velocity given by
\cite{donnelly2}:-
\begin{displaymath}
v_i = \frac{\kappa}{4 \pi R} \Biggl[ \ln \biggl (\frac{8R}{a} \biggr)
- \frac{1}{2} \Biggr]
\end{displaymath}
where $\kappa = h / m \approx 10^{-7} {\rm m}^{2} {\rm s}^{-1}$, $R$ is
the radius of a loop of string and $a$ is the vortex core parameter
first introduced by Feynman.
When the decompression occurs, there will be superflows within the
bellows. While the walls may perhaps not be terribly effective in
generating these, it is guaranteed that superfluid will get forced out
of the end of the filling capillary producing a superflow directly
through the middle of the thermal vortex population. One can estimate
$v_s$, the velocity of the superflow out of the capillary from the change in
pressure, $\Delta P$ and the density of the liquid helium inside the
capillary.
\begin{displaymath}
V \Delta P = \frac{1}{2} (\rho V) v_s^2.
\end{displaymath}
For the change in pressure of about $10^6$Pa actually used this gives a
velocity of roughly the critical superflow or slightly below, so it is
possible that vorticity is generated by supercritical flows. However,
since vorticity is not generated unless the quenches are sufficiently
close to the transition, it is safe to assume that supercritical flows
alone are not responsible for vortex formation and that the only
possible explanation for the vorticity formed during quenches through
the transition is that already described.

Even if the flow is not supercritical, however, the effect of a superflow on
a population of vortex loops is to expand loops greater than a certain size
with an appropriate velocity relative to the superflow according to
\begin{displaymath}
\frac{dR}{dt} = \Biggl [
\frac{ \gamma } { \rho_s \kappa }
\Biggr ]
\Biggl  ( v_s - v_i \Biggr ).
\end{displaymath}
Thus loops with sufficiently large radii that their velocity is less
than the velocity of the superflow and are also oriented so that some
component of their velocity is parallel to the superflow, or roughly
half the loops if the distribution is isotropic, will grow in the flow.
This could result in the magnification of the thermal vortex line density.

For a thermal population of vortex loops, however, one expects almost
all of the loops to be of a size roughly equal to the thermal coherence
length. Sufficiently far from the transition, the vortex core parameter $a$,
equal to the thermal coherence length is only of the order of $10^{-8}$m and
there will be almost no loops sufficiently large, and therefore sufficiently
slow in comparison with the superflow, to be expanded. Near the
transition, however, the coherence length grows increasingly large and thus
the velocity of the typical coherence sized loop becomes smaller and
smaller as one approaches the phase transition. Sufficiently close to
the transition, even coherence sized loops will have velocities slower
than the superflow and will be expanded.

Assuming that the number density of loops of a given length decays
fairly steeply with length, like $l^{-5 / 2}$ or exponentially for example,
it is a reasonable approximation to say that almost all of the string is
in coherence length sized loops. Also, the density of thermal vortices is
only significant near the transition. Thus, there will be almost no
amplification of the thermal density by the superflows in the fluid
unless the initial sub-critical temperature is such that
coherence sized loops move slower than the superflow velocity or in other
words
\begin{displaymath}
\frac{\kappa}{4 \pi \xi(T)} \biggl ( \ln (8) - 0.5 \biggr ) < \sqrt
{ \frac{ \Delta P }{\rho} },
\end{displaymath}
or alternatively
\begin{displaymath}
\frac{T}{T_c} > 1 - \Biggl ( \frac{ 4 \pi v_s \xi_0 }
{\kappa [ \ln 8 - 0.5 ] } \Biggr ) ^2.
\end{displaymath}
In this scenario one would expect a fairly sharp temperature above which
quenches start to produce significant vorticity as opposed to the simple
stirring scenario in which one expects a gradual logarithmic increase in
the vorticity. Also, the fractional increase in the vortex density
or fractional amplification in this case is roughly
\begin{displaymath}
\frac{\Delta n}{n}
\approx
\frac{1}{R} \, \frac{dR}{dt} \, \Delta t
\approx
\frac { B(v_s - v_i) }{R} \Delta t
\approx
\frac{ \Delta t v_s \epsilon^{1/2} } {\xi_0}
\Biggl [ 1 - \frac{ \kappa \epsilon^{1/2} } { 10 \xi_0 v_s } \Biggr ]
\end{displaymath}
where $\epsilon$ is the relative temperature $1-T/T_c$ and $B \approx \gamma
/ \rho \kappa$ is a dimensionless parameter of order 6 in the region of
interest. Between $\epsilon \approx 10^{-3}$ and {$\epsilon \approx
10^{-4}$, this amplification factor changes by only a factor of
$\sqrt{10}$, being of order $10^6$ or $10^7$. Using both experimental and
renormalisation group improved values, the thermal seed density varies only
slowly in the regime of interest, although the density does vary
by a factor of roughly $10^5$ if mean field critical exponents are used.
It might therefore be expected that one would see a well defined temperature
at which the amplification switches on.

In addition, if ILF is the dominant process, since conformal
invariance demands that there
are no vortices actually at the transition, as is consistent with the
mean field calculation above (\ref{napp}), there ought also to be an initial
temperature
above which no vorticity is observed following the pressure quench.
However, this temperature may turn out to be so close to the critical
temperature as to be unobservable.

\section{Summary And Conclusions}

We have used an equilibrium finite temperature field theoretical description
of superfluid $^4$He to predict the thermal density of vortex loops in the
superfluid phase. This has been used in an adiabatic approximation to predict
the vortex density produced in the Lancaster experiment for quenches from
below the $\lambda$-transition.
Although mean field theory is a very poor approximation in the region of
interest, it is unlikely to be inaccurate by more than a few orders of
magnitude at most.
Thus the fact that a mean field theory estimate
of the thermal vortex density below the transition gives a value several
orders of magnitude below that observed implies that the vorticity observed
is not simply the thermal population. Either the thermal population is
amplified in some way or there is some other significant form of vorticity
production. Here we have argued that the thermal density could be amplified
by a superflow produced from the capillary which is used to fill the
experimental cell. This is entirely analogous to the intrinsic thermal
nucleation of defects known as the ILF process. The experimental signature of
this scenario would be a sharp cut-off temperature such that quenches starting
below this produce too little vorticity to be observed. There should
also be an upper cut-off although this might be impossible to observe.

There are, however, problems. Firstly although it is difficult to come up
with an alternative scenario with such copious vortex production, there are
experimental problems which make it difficult to entirely eliminate other
methods of vortex production. Secondly there is the theoretical problem that
even if mean field theory were a good approximation, the field theoretical
description of $^4$He is inaccurate since the coherence length is never very
much greater than the interatomic spacing.


\section*{Acknowledgements}

The authors would like to thank  R.A.M. Lee, P.V.E. McClintock, E. Mottola,
G. Pickett, R.J. Rivers and W.H. Zurek, for useful conversations, particularly
R.J. Rivers for his understanding of the method of Wiegel.  We also
acknowledge the hospitality of the Isaac Newton Institute for
Mathematical Sciences, Cambridge and of the Low Temperature Laboratory,
Helsinki University of Technology, where some of this work was performed.







\end{document}